\documentclass[11pt]{article}
\usepackage{amssymb}
\usepackage{amsmath}
\usepackage{latexsym}

\catcode`@=11
\renewcommand{\section}{\@startsection{section}{1}{0pt}{\medskipamount}
{\medskipamount}{\large\bf}}
\numberwithin{equation}{section}
\catcode`@=12
\def\th{\theta}
\def\a{\alpha}
\def\b{\beta}
\def\g{\gamma}
\def\de{\delta}
\def\e{\epsilon}
\def\l{\lambda}

\def\ve{\varepsilon}
\def\vp{\varphi}
\def\p{\phi}
\def\s{\sigma}

\def\sfrac#1#2{{\textstyle\frac{#1}{#2}}}
\def\m{\mu}
\def\n{\nu}

\def\pa{\partial}

\def\rd#1{\buildrel{_{_{\hskip 0.01in}\rightarrow}}\over{#1}}
\def\ld#1{\buildrel{_{_{\hskip 0.01in}\leftarrow}}\over{#1}}
\newcommand{\im}{\mathrm{i}}
\newcommand{\sh}{\mathrm{sh}}
\newcommand{\ch}{\mathrm{ch}}
\newcommand{\cth}{\mathrm{cth}}
\newcommand{\diff}{\mathrm{d}}

\newcommand{\zb}{{\bar{z}}}
\newcommand{\yb}{{\bar{y}}}
\newcommand{\td}{\dot{T}}
\newcommand{\rc}{{\mathbb{R}^4}}
\newcommand{\rct}{{\mathbb{R}^4_\theta}}
\newcommand{\R}{{\mathbb{R}}}
\newcommand{\Id}{\mathbf{1}_2}

\newcommand{\Hcal}{{\cal H}}
\newcommand{\fh}{\hat{f}}
\newcommand{\gh}{\hat{g}}

\def\>{\rangle}
\def\<{\langle}
\def\+{\dagger}

\begin{document}
\begin{titlepage}
\setcounter{page}{0}
\begin{flushright}
hep-th/0209153
\end{flushright}

\vskip 2.5cm

\begin{center}

{\Large\bf On Noncommutative Merons and Instantons}

\vspace{15mm}

{\large Filip Franco-Sollova$^*$ \ and \ Tatiana A. Ivanova~$^\+$}
\\[5mm]
\noindent ${}^*${\em Institut f\"ur Theoretische Physik,
Universit\"at Hannover \\
Appelstra\ss{}e 2, 30167 Hannover, Germany }\\
{Email: filip@itp.uni-hannover.de}
\\[5mm]
\noindent ${}^\+${\em Bogoliubov Laboratory of Theoretical Physics, JINR\\
141980 Dubna, Moscow Region, Russia}\\
{Email: ita@thsun1.jinr.ru}

\vspace{1.5cm}

\begin{abstract}
The Yang-Mills (YM) and self-dual Yang-Mills (SDYM) equations 
on the noncommutative Euclidean four-dimensional space are 
considered.  We introduce an ansatz for a gauge potential reducing 
the noncommutative SDYM equations to a difference form of the Nahm 
equations. By constructing solutions to the difference Nahm equations, 
we obtain solutions of the noncommutative SDYM equations. They are 
noncommutative generalizations of the known solutions to the SDYM 
equations such as the Minkowski solution, the one-instanton solution 
and others.  Using the noncommutative deformation of the 
Corrigan-Fairlie-'t~Hooft-Wilzek ansatz, we reduce the noncommutative 
YM equations to equations on a scalar field which have meron solutions 
in the commutative limit and show that they  have no such solutions in 
the noncommutative case. To overcome this difficulty, another ansatz 
reducing the  noncommutative YM equations to a system of difference 
equations on matrix-valued functions is used. For self-dual configurations 
this system is reduced to the difference Nahm equations. 

\end{abstract}

\end{center}
\end{titlepage}

\section{Introduction}

There is a continued interest in  noncommutative 
field theories since many of them appear in a certain zero-slope
limit of open strings coupled to a $B$-field background~\cite{Seiberg:1999}.
Noncommutativity of coordinates gives an opportunity
to introduce nonlocality into field theory without losing control 
over its structure. Before attempting to quantize noncommutative 
field theories, it is desirable to characterize the moduli space of 
their classical configurations. The generalizations of classical 
solutions to the noncommutative case began in 1998 when Nekrasov and 
Schwarz~\cite{NS} gave the first examples of noncommutative instantons. 
Nonperturbative soliton-like solutions also attracted much attention, 
since they admit a $D$-brane interpretation in the context of string 
theory. The consideration of noncommutative solitons was initiated by 
Gopakumar, Minwalla and Strominger~\cite{Gopakumar:2000}. Since then, 
a lot of papers on noncommutative instantons and solitons in various 
field theories has appeared (see e.g.~\cite{Fur}\,-\,\cite{Ishikawa}); 
it is impossible to mention all of them. For more references see 
review papers~\cite{Nekrasov}\,-\,\cite{Konechny:2001}.

In this paper we consider the  noncommutative Yang-Mills (ncYM) and 
noncommutative self-dual Yang-Mills (ncSDYM) equations in four dimensions 
and discuss the noncommutative generalizations~\cite{Correa1, LP1} of the 
Belavin-Polyakov-Schwarz-Tyupkin (BPST) and Corrigan-Fairlie-'t~Hooft-Wilzek 
(CFtHW) ans\"atze for $U(2)$ instantons. We generalize the noncommutative 
BPST-like ansatz~\cite{Correa1} to the case of $U(n)$ group and  reduce the 
ncSDYM equations to a difference form of Nahm's equations.  Our ansatz is a 
noncommutative generalization of the ansatz from~\cite{Ivanova:tu} 
where the BPST ansatz~\cite{Belavin} was extended to the $SU(n)$ gauge group 
and some solutions to the SDYM equations were constructed using the solutions 
of Nahm's equations~\cite{Nahm}. Solving the difference Nahm equations permits 
us to obtain some explicit solutions of the ncSDYM equations. Among them there 
are  noncommutative generalizations of Minkowski's~\cite{Mink}, 
BPST instanton~\cite{Belavin} and abelian 't~Hooft~\cite{'tHooft} solutions.
We show that the same ansatz reduces the ncYM equations to a system of 
difference equations which in the commutative limit have meron solutions.

\medskip

\section{Noncommutative YM and SDYM equations}

{\bf  YM theory on commutative space.}
We consider the Euclidean space $\rc$ with the metric $\de_{\m\n}$, 
a gauge potential $A=A_{\mu}(x)\diff x^\mu$ and the Yang-Mills 
field $F=\diff A+A\wedge A$ with the components
 $F_{\mu\nu}=\pa_{\mu}A_{\nu}-\pa_{\nu}A_{\mu}+[A_{\mu},A_{\nu}]$,
where $x=(x^\m)\in \rc$, $\pa_{\mu} :=\pa /\pa x^\mu$ and $\mu,\nu,
\ldots=1,2,3,4$. Both 
$A_{\mu}$ and $F_{\mu\nu}$ take values in the Lie algebra $u(n)$.
The YM equations in $\rc$ have the form
\begin{equation}\label{ym1}
D_\mu F_{\mu\nu}\ =\ 0\ ,
\end{equation}
where $D_\mu := \pa_\mu + {\rm ad}\, A_\mu$. 
The SDYM equations are~\cite{Belavin}
\begin{equation}\label{sdym1}
 F_{\mu\nu}\ =\ \sfrac{1}{2}\ve_{\mu\nu\rho\s}F_{\rho\s} \ ,
\end{equation}
where $\ve_{\mu\nu\rho\s}$ is the completely antisymmetric tensor in 
$\rc$, with $\ve_{1234}=1$. By virtue of the Bianchi identity 
$\ve_{\mu\nu\rho\s}D_\nu F_{\rho\s}  \ =0$,
 each solution of the SDYM equations (\ref{sdym1}) satisfies the YM equations  
(\ref{ym1}).  

{}For a set of complex coordinates in $\rc$,
\begin{equation}
y=x^1+\im x^2\ , \quad z=x^3-\im x^4\ , \quad \yb=x^1-\im x^2\ , \quad 
\zb=x^3+\im x^4\ ,
\label{ccrf}
\end{equation}
the corresponding components of a gauge potential read
\begin{equation}
A_y=\sfrac{1}{2}(A_1-\im A_2)\ , \quad A_z=\sfrac{1}{2}(A_3+\im A_4)\ , \quad
A_{\yb}=\sfrac{1}{2}(A_1+\im A_2)\ , \quad A_{\zb}=\sfrac{1}{2}(A_3 -\im A_4)\ , 
\end{equation}
and the SDYM equations take the form
\begin{equation}
F_{yz}=0\ , \quad F_{\yb \, \zb}=0 \quad\mbox{ and }\quad 
F_{y \yb}+F_{z \zb}=0 \ .
\label{cym}
\end{equation}

\medskip

{\bf Noncommutative setting.}
In comparison with the  ordinary field theory, in noncommutative field theories 
(see e.g.~\cite{Douglas:2001}) the standard (commutative)  product of functions is replaced 
by the noncommutative star product,
\begin{equation}{\label{star}}
(f \star g)(x)\ =\ f(x)\,\exp\,\left\{ \frac{\im}{2}
{\ld{\partial}}_\mu \,\theta^{\mu\nu}\, {\rd{\partial}}_\nu \right\}\,g(x) \ ,
\end{equation}
with a constant antisymmetric tensor~$\th^{\mu\nu}$.
In the star-product formulation, the components of the noncommutative field 
strength have the form
\begin{equation}
{}F_{\mu\nu}\ =\ \pa_\mu A_\nu-\pa_\nu A_\mu+A_\mu\star A_\nu-A_\nu\star A_\mu\ ,
\end{equation}
and the YM and SDYM equations look formally to be unchanged.
However, the nonlocality of the star product makes explicit calculations
too cumbersome. Therefore, one usually uses the Moyal-Weyl map
between ordinary commutative functions $f, g$ from (\ref{star})
and operators $\hat f, \hat g$ acting in the two-oscillator Fock space.

To be more precise, let us first introduce a four-dimensional noncommutative 
Euclidean space $\rct$ as a space the coordinates $\hat{x}^\mu$ of which 
satisfy $[\hat{x}^\mu,\hat{x}^\nu]=\im\theta^{\mu\nu}$. By a change of 
coordinates we can transform the tensor $\theta^{\mu\nu}$ to the form
 \begin{equation}\label{th}
\theta^{12}= -\theta^{21}=\e\theta^{34}=-\e\theta^{43}=\th >0\ ,
\end{equation}
where $\e =1$ for the self-dual and $\e = -1$ for the anti-self-dual tensor
$\th^{\mu\nu}$.
Then, in complex coordinates (\ref{ccrf}) the choice (\ref{th}) leads to
\begin{equation}\label{He}
[\hat y,\hat \yb]=2\th\ ,  \quad [\hat z, \hat \zb]=-2\e\th\ ,
\end{equation}
and other commutators are zero.
The coordinate derivatives are now inner derivations of the algebra of functions, i.e.
\begin{equation}\label{der}
\hat\pa_y \hat f = -\frac{1}{2\th} [\hat\yb,\hat f]\ , 
\quad \hat\pa_z \hat f =\frac{1}{2\e\th}[\hat\zb,\hat f]\ ,
\quad
\hat \pa_{\yb} \hat f = \frac{1}{2\th} [\hat y,\hat f]\ , 
\quad \hat\pa_{\zb} \hat f =-\frac{1}{2\e\th}[\hat z,\hat f]\ ,
\end{equation}
where $\hat f$ is any function of $\hat y, \hat \yb, \hat z, \hat \zb$.
The obvious representation space for the Heisenberg algebra~(\ref{He})
is the two-oscillator Fock space~$\Hcal$ spanned by 
$\{ |n_1,n_2\>\ \textrm{with}\ n_1,n_2=0,1,2,\ldots\}$.
In $\Hcal$ one can introduce an integer ordering of states~\cite{Rangamani:2001cn},
\begin{equation} \label{fock}
|k\>\ =\ |n_1,n_2\> \ =\
\frac{\hat\yb^{n_1}\,\hat\zb_\e^{n_2}\,|0,0\>}
{\sqrt{n_1!n_2!(2\th)^{n_1+n_2}}}
\qquad\textrm{with}\qquad
\hat{z}_\e\ :=\ 
\sfrac{1-\e}{2}\,\hat{z}\ +\ \sfrac{1+\e}{2}\,\hat{\bar{z}}
\end{equation}
and $k=n_1+\sfrac{1}{2}(n_1+n_2)(n_1+n_2+1)$.
The coordinates, functions of them  and all fields  are  regarded as 
operators in~$\Hcal$. The Moyal-Weyl map gives the operator equivalent 
of star multiplication and integration, i.e.
\begin{equation} \label{trace}
\textrm{if}\quad f\mapsto\fh\ , \quad  g\mapsto\gh\  \quad \textrm{then}\quad
f\star g\ \mapsto\ \fh\,\gh \qquad\textrm{and}\quad
\int\! \diff^4{x}\,f\ =\
(2\pi \theta)^2 \,\mbox{Tr}_\Hcal\, \fh\ ,
\end{equation} 
where `$\mbox{Tr}_\Hcal$' denotes the trace over the Fock space~$\Hcal$. For
explicit formulae of this correspondence see 
e.g.~\cite{Nekrasov}\,-\,\cite{Konechny:2001}.

Let us introduce the following operators acting in the Fock space~$\Hcal$:
\begin{equation} \label{X}
X_\mu \ :=\ \hat A_\mu + \im\th_{\mu\nu} \hat x^\nu\ ,
\end{equation}
where $\hat A_\mu$ are operators in $\Hcal$ corresponding to the 
components $A_\mu$ of a
gauge potential $A$ in $\rc$, $\th_{\mu\s}\th^{\s\nu}=\de_{\mu}^{\nu}$.
In terms of $X_\mu$ the gauge field strength components are
\begin{equation} \label{ncF} 
\hat F_{\mu\nu}= [X_\mu ,X_\nu ] - \im \th_{\mu\nu}\ . 
\end{equation} 
In the operator formulation the noncommutative version of the YM equations 
(\ref{ym1}) reads
\begin{equation} \label{ncym1}
[X_\mu , [X_\mu , X_\nu ]]\ =\ 0\ .
\end{equation}
The ncSDYM equations have the form
\begin{equation} \label{ncsdym1}
[X_\mu , X_\nu ]\ =\ \sfrac{1}{2}\ve_{\mu\nu\rho\s} [X_\rho , X_\s ] + 
\im(\th_{\m\n} - \sfrac{1}{2}\ve_{\mu\nu\rho\s}\th_{\rho\s} )  \ . 
\end{equation}
For $\th^{\m\n}$ from (\ref{th}) the above equations in complex coordinates 
are equivalent to
\begin{equation} \label{ncsdym2}
[X_{y}, X_{z} ]=0\ , \quad
[X_{\yb}, X_{\zb} ]=0\ , \quad  [X_{y}, X_{\yb}]+ [X_{z}, X_{\zb} ] + \frac{\e -1}{2\th}=0\ .
\end{equation}
For convenience we shall omit the hats over the operators.

\medskip

\section{Noncommutative SDYM and difference Nahm equations}

{\bf Noncommutative generalization of the CFtHW ansatz.}
In the commutative space $\rc$, the Corrigan-Fairlie-'t~Hooft-Wilczek (CFtHW) 
ansatz for a gauge potential has the form
\begin{equation}\label{cfthw}
A_\mu\ =\  - \bar\eta^a_{\mu\nu}\frac{\s_a}{2\im}
\p^{-1}\pa_\nu\p \ ,
\end{equation}
where
\begin{equation}
\bar\eta^a_{bc}=\e^a_{bc}\ ,\quad  
\bar\eta^a_{\m 4}\ =-\de^a_\mu\ , \quad 
\bar\eta^a_{\m \n}=-\bar\eta^a_{\n \m}
\end{equation}
is the anti-self-dual 't~Hooft tensor~\cite{Prasad:1980yy},
$\ \mu , \nu ,... =1,...,4$, $\ a,b,c=1,2,3$ ,
and $\sigma_a$ are 
the Pauli matrices,
\begin{equation}
\sigma_1=
\left( \begin{array}{rr} %
      0 & 1  \\ 1 & 0  
\end{array}\right), \; \;
\sigma_2=
\left( \begin{array}{rr} %
        0 & -\im  \\ \im & 0  
\end{array}\right), \; \;
\sigma_3=
\left( \begin{array}{rr} %
        1 & 0  \\ 0 & -1 
\end{array}\right)\ .
\label{pauli}
\end{equation}
The substitution of (\ref{cfthw}) into the SDYM equations (\ref{sdym1})
reduces them to the equation $\p^{-1}\Box\p =0$.
Taking the solution of the Laplace equation
\begin{equation}\label{laplsol}
\phi_n = 1+\sum^n_{k=1}\frac{\Lambda^2_k}{r^2_k}\ ,
\end{equation}
where $r^2_k:=(x_\m-a^k_\m)(x_\m-a^k_\m)$ and $\Lambda_k$ are constants,  
one obtains a singular multi-instanton 't~Hooft solution of the SDYM 
equations (\ref{sdym1}).

The noncommutative generalization of the CFtHW ansatz (\ref{cfthw}) was 
introduced\footnote{The nonreal operator form of (\ref{cfthw}) was earlier 
considered in~\cite{NS}.} 
in~\cite{Correa1} and derived via the twistor approach in~\cite{LP1}.
It reads
\begin{equation}\label{Amu}
A_\mu\ =\ \bar\eta^a_{\mu\nu}\frac{\s_a}{2\im}
\left(\p^{\frac{1}{2}}\pa_\nu\p^{-\frac{1}{2}}-\p^{-\frac{1}{2}}\pa_\nu\p^{\frac{1}{2}}\right)
+ \frac{{\mathbf 1}_2}{2}
\left(\p^{-\frac{1}{2}}\pa_\mu\p^{\frac{1}{2}}+\p^{\frac{1}{2}}\pa_\mu\p^{-\frac{1}{2}}\right)
\end{equation}
and reduces the ncSDYM equations to the equation
\begin{equation}\label{redop}
\phi^{-\frac{1}{2}}
(\pa_y\pa_{\bar y}\phi + \pa_z\pa_{\bar z}\phi )\,\phi^{-\frac{1}{2}}\ =\ 0\ .
\end{equation}
It is reasonable to assume that a solution $\phi_n$ of 
this equation looks as the standard 't~Hooft  solution (\ref{laplsol})
producing $n$-instantons, but the direct substitution of such $\phi_n$ into 
(\ref{Amu}) yields a gauge field which is not self-dual on an 
$n$-dimensional subspace of the Fock space. This deficiency was 
analyzed in~\cite{Correa1} for the $n=1$ case and in~\cite{LP1} for the case of 
arbitrary $n$. In~\cite{LP1} it was proposed to use  
a suitable Murray-von Neumann transformation after a specific projection of the gauge 
potential. The proper noncommutative 't~Hooft multi-instanton field 
strength was given explicitly~\cite{LP1}, but its gauge 
potential was not obtained in the explicit form.  
 
\medskip

{\bf Noncommutative generalization of the BPST ansatz.}
The authors of \cite{Correa1} discussed another possibility of getting over the 
difficulties connected with the naive extension of the CFtHW ansatz. Namely, 
they suggested using the BPST ansatz~\cite{Belavin}  
for construction of noncommutative $U(2)$ self-dual field configurations. They have got
them explicitly, but the reality of a self-dual gauge potential and the YM field
was lost. It seems that there exists a complex gauge transformation
of their solution to a real form, since the Lagrangian evaluated on the solution is
real and the topological charge $Q$ is equal to one.

Here we want to generalize the BPST-like ansatz of \cite{Correa1} in the way
it was done in the commutative case~\cite{Ivanova:tu}. There,  $su(n)$-valued
components of a gauge potential in $\rc$ were chosen in the form  
\begin{equation}\label{ita}
A_\m = 2\eta^a_{\m\n} x_\n T_a (\tau ) + 2 x_\m T_4 (\tau )\ ,
\end{equation}
where $\tau := x_\m x^\m$,  $T_\m(\tau)$ are $su(n)$-valued functions of $\tau$
and $\eta^a_{\m\n}$ are the components of the self-dual 't~Hooft tensor~\cite{Prasad:1980yy},
\begin{equation}
\eta^a_{bc}=\e^a_{bc}\ ,\quad  \eta^a_{\m4}=\de^a_\m\ ,\quad  
\eta^a_{\m\n}=-\eta^a_{\n\m}\ .
\end{equation} 
A direct substitution of (\ref{ita}) into the SDYM equations (\ref{sdym1})
reduces them to Nahm's equations
\begin{equation}\label{nahm}
\td_1 =-[T_2,T_3]-[T_4,T_1]\ , \quad 
\td_2 =-[T_3,T_1]-[T_4,T_2]\ ,  \quad
\td_3 =-[T_1,T_2]-[T_4,T_3]\ ,
\end{equation}
where the dot over $T_\m$ denotes the derivative w.r.t. $\tau$, $\td_\m :=\frac{dT_\m}{d\tau}$.  
So, in  \cite{Ivanova:tu} it was shown that to any  solution of Nahm's equations 
(\ref{nahm}) there corresponds the explicit solution (\ref{ita}) 
of the SDYM equations (\ref{sdym1}). We shall generalize 
the ansatz (\ref{ita}) to noncommutative $U(n)$ gauge theory in the self-dual space $\rct$.

{}For the complex coordinates in the self-dual (i.e. $\e =1$) noncommutative Euclidean 
space $\rct$ we have 
$$[y,\yb ]=2\th\ , \quad [z,\zb ]=-2\th\ .$$ 
Let us consider the noncommutative generalization of 
the ansatz (\ref{ita})
$$A_y = - 2\im\bar z T_y(\tau) + 2\im \bar y T_{\bar z}(\tau)\ , 
\quad
A_z = - 2\im \bar y T_{\bar y}(\tau )-2\im \bar z T_z(\tau)\ , $$
\begin{equation}
A_{\bar y} = 2\im z {T}_{\yb}(\tau)  - 2\im  y{T}_{z}(\tau ) \ ,
\quad
A_{\bar z} = 2\im  y{T}_{y}(\tau ) + 2\im  z{T}_{\zb}(\tau ) \ , 
\label{ans1}
\end{equation}
where $T_\m$ are some $u(n)$-valued functions of the operator 
$\tau :=\bar y y+ \bar z z$. Notice that for ${T}_{\m}(\tau )$ in 
$\rct$ we have
\begin{equation}\label{tyyt}
T_\m y=yT^-_\m\ ,\quad T_\m \zb = \zb T^-_\m\ ,\quad T_\m \yb =\yb T^+_\m\ ,\quad
T_\m z=zT^+_\m\ ,
\end{equation}
where $T^\pm_\m := {T}_{\m}(\tau \pm 2\th)$.
The ansatz (\ref{ans1}) gives us a $u(n)$-valued gauge potential $A=A_\m^aJ_a$ with 
complex components $A_\m^a$, since due to (\ref{tyyt}) we have
$A^\+ = (A_\m^a)^\+ J_a^\+= - (A_\m^a)^\+ J_a\ne -A$. Here
 $J_a$ are the generators of $U(n)$. 

\medskip

{\bf Reduction to  difference Nahm equations.}
Let us introduce new variables
\begin{equation}\label{nnnn}
N_y := 2\im T_y\ ,\quad N_\yb := 2\im T_\yb \ ,
\quad N_z := 2\im T_z - \frac{1}{2\th}   \ ,\quad 
 N_\zb := 2\im T_\zb - \frac{1}{2\th}   \ ,
\end{equation}
in which  the operators $X_\m$ from (\ref{X}) for the ansatz (\ref{ans1})
read
\begin{equation}\label{ans2}
X_y = -\zb N_y + \yb N_\zb\ ,\quad X_z =  -\yb N_\yb - \zb N_z\ ,\quad
X_\yb =  z N_\yb - y N_z \ ,\quad X_\zb = y N_y + z N_\zb \ .
\end{equation}
After some computations we obtain
$$
[X_{y}, X_{z} ]\ =\ \zb^2Q_1 + \yb^2 Q_2 + \yb\zb Q_3 \ , 
$$
$$
[X_{\yb}, X_{\zb} ]\ =\ y^2Q_1 + z^2 Q_2 - yz Q_3 \ ,
$$ 
\begin{equation}\label{ncsdym3} 
[X_{y}, X_{\yb}]+ [X_{z}, X_{\zb} ]\ = \   2y\zb Q_1 - 2z\yb Q_2 + (\yb y - z\zb ) Q_3 \ ,
\end{equation}
where
\begin{equation}\label{qqq}
Q_1:= N_y^-N_z - N_z^-N_y \ , \quad
Q_2:= N_\yb^+N_\zb - N_\zb^+N_\yb \ , \quad
Q_3:= N_y^+N_\yb - N_\yb^-N_y + N_z^+N_\zb -  N_\zb^- N_z  \ .
\end{equation}
Recall that $N_\mu^\pm := {N}_{\m}(\tau \pm 2\th)$.
So, a direct substitution of (\ref{ans2}) into the ncSDYM equations
(\ref{ncsdym2}) in self-dual $\rct$ ($\e =1$) leads to 3 equations 
\begin{equation}\label{ncnahm}
Q_1=0\ ,\quad Q_2=0\ ,\quad Q_3=0\ ,
\end{equation}
or, in terms of complex $u(n)$-valued functions $N_\m$,
\begin{equation}\label{ncnahm1}
N_y^-N_z - N_z^-N_y=0 \ , \quad
N_\yb^+N_\zb - N_\zb^+N_\yb =0 \ , \quad
N_y^+N_\yb - N_\yb^-N_y + N_z^+N_\zb -  N_\zb^- N_z=0.  
\end{equation}
Equations (\ref{ncnahm1}) are similar (but not coincident\footnote{
There may exist different nonequivalent discretizations of differential 
equations. See e.g. discussion in~\cite{bob} of nonequivalent 
discretizations of the sine-Gordon equation.}) 
to the discrete Nahm equations introduced in~\cite{BA} and discussed in the context 
of hyperbolic monopoles in~\cite{MS, Ward}.

In terms of $T_\m (\tau )$,  eqs.(\ref{ncnahm1}) read
$$
\frac{1}{2\theta}\{ T_{\bar y}(\tau +2\theta) - T_{\bar y}(\tau) \}  =
 2\im  \{T_{\bar y}(\tau +2\theta )T_{\bar z}(\tau ) -
 T_{\bar z}(\tau + 2\theta ) T_{\bar y}(\tau )\}\ ,
$$
\begin{equation}
\frac{1}{2\theta}\{ T_y(\tau +2\theta ) - T_y(\tau ) \}  =
 2\im  \{T_z(\tau ) T_y(\tau +2\theta ) - T_y(\tau ) T_z(\tau +2\theta )\}\ ,
\label{neq}
\end{equation}
$$
\frac{1}{2\theta}\{ T_z(\tau +2\theta ) - T_z(\tau )  + T_{\bar z}(\tau ) -
 T_{\bar z}(\tau -2\theta ) \} =$$
$$
= 2\im \{ T_y(\tau +2\theta )T_{\bar y}(\tau ) - T_{\bar y}(\tau -2\theta )
T_y(\tau ) + T_z(\tau +2\theta )T_{\bar z}(\tau ) - 
T_{\bar z}(\tau -2\theta )T_z(\tau )\}\ .
$$
We shall call equations (\ref{neq}) {\it the difference Nahm equations}.
Thus, the ansatz  (\ref{ans1}) reduces
the noncommutative SDYM equations in the self-dual Euclidean
space $\rct$ to the difference Nahm equations (\ref{neq}).
So, to each solution of eqs.(\ref{neq}) one may correspond a solution of the 
ncSDYM equations (\ref{ncsdym2}).

In the limit $\theta \rightarrow 0$, eqs.(\ref{neq}) take the form
\begin{equation}
\td_{\yb}=2\im [T_{\yb},T_{\zb}]\ , \quad
\td_y=2\im [T_z,T_y] \ , \quad
\td_z+\td_{\zb}=2\im \left\{[T_z,T_{\zb}]+[T_y,T_{\yb}]\right\}\ .
\label{nah}
\end{equation}
If one introduces
$$
T_1=T_y+ T_\yb\ ,\quad  T_2=\im (T_y - T_\yb)\ , \quad 
T_3=T_z+T_\zb \ ,\quad  T_{4}=\im (T_\zb - T_z)\ ,
$$
then one obtains differential Nahm's equations (\ref{nahm}).
So, in the commutative limit $\theta \to 0$, the result of our reduction 
agrees with the one from \cite{Ivanova:tu}. 

\medskip

{\bf Solutions of difference Nahm's equations.} 
Consider $U(2)$ as a gauge group. Its generators have the form
$J_a=\sigma_a/2\im$, $J_4=\im {\Id}/{2}$.
Introduce the matrices
\begin{equation}
J_y:=\frac{1}{2}(J_1-\im J_2)=
\frac{1}{2\im}\begin{pmatrix}0&0\\1&0\end{pmatrix}\ ,
\quad
J_{\bar y}:=\frac{1}{2}(J_1+\im J_2)=
\frac{1}{2\im}\begin{pmatrix}0&1\\0&0\end{pmatrix}\ ,
\nonumber
\end{equation}
\begin{equation}
J_z:=\frac{1}{2}(J_3+ J_4)=-\frac{1}{2\im}\begin{pmatrix}0&0\\0&1
\end{pmatrix}\ ,\quad
J_{\bar z}:=\frac{1}{2}(J_3- J_4)=\frac{1}{2\im}\begin{pmatrix}1&0\\0&0
\end{pmatrix}\ .
\label{gen}
\end{equation}
Calculations give
\begin{align}
J_{\bar y}J_{\bar z}&=J_yJ_z=J_{\bar z}J_{y}=J_zJ_{\bar y}=J_zJ_{\bar z}=J_{\bar z}J_z=0\ , 
\quad J_{\bar z}J_{\bar y}= - J_{\bar y}J_z=\frac{1}{2\im} J_{\bar y}\ , 
\nonumber\\[8pt]
J_zJ_y &= - J_yJ_{\bar z}=   -\frac{1}{2\im} J_y\ ,\quad
J_yJ_{\bar y}=J_zJ_z= -\frac{1}{2\im}J_z\ ,
\quad  J_{\bar y}J_y= J_\zb J_\zb=\frac{1}{2\im}J_{\bar z}\ .
\label{prop}
\end{align}

In searching for solutions to the difference Nahm equations (\ref{neq})
we consider the following ansatz for $T_\m (\vp)$: 
\begin{equation}
T_y= \frac{\g}{\sh(\vp )}J_y\ , \quad
T_{\bar y}= \frac{\g}{\sh(\vp )} J_{\bar y} \ ,\quad
T_z= f_z(\vp )J_z\ , \quad
T_{\bar z}=f_{\bar z}(\vp )J_{\bar z}\ ,
\label{ans3}
\end{equation}
where $\vp: = \a (r^2 + \b^2)$, $r^2\equiv \tau :=\bar yy +\bar zz$  \ and\ 
$\a$, $\b$, $\g$ are some parameters. In this subsection we use the notation
$r^2$ to have more similarity with solutions of the commutative theory in $\rc$. 
Note that in $\rct$ for any function $f(\vp (r^2))$ we have 
$f(\vp (r^2 + 2\th))=f(\vp (r^2) + 2\a\th)$.

Substituting (\ref{ans3}) into the first two equations of (\ref{neq}) and using 
(\ref{prop}), we obtain
\begin{equation}
f_z=\frac{1}{2\th }(\ch (2\a\th) -1 + \sh(2\a\th ) \cth(\vp ))\ , 
\quad
f_{\bar z}=\frac{1}{2\th }(1 - \ch(2\a\th ) + \sh(2\a\th )\cth(\vp ) )
\ .
\label{fz}
\end{equation}
To find the value of the parameter $\g$, we substitute (\ref{ans3}) with (\ref{fz}) 
into the last equation of (\ref{neq}) and obtain
$$
\g =\frac{\sh(2\a\th )}{2\th}\ .
$$
So, the solution of the difference Nahm equations (\ref{neq})
has the form
$$
T_y= \frac{\sh(2\a\th )}{2\th\ \sh(\a (r^2 + \b^2) )}J_y\ , \quad
T_{\bar y}=\frac{\sh(2\a\th )}{2\th\ \sh(\a (r^2 +\b^2 ))} J_{\bar y} \ ,
$$
$$
T_z=\frac{1}{2\th }\left\{\ch (2\a\th)-1 +  \sh(2\a\th )
\cth(\a (r^2 +\b^2 ))\right \}J_z\ ,
$$
\begin{equation}
T_{\bar z}=\frac{1}{2\th }\left\{1 - \ch(2\a\th ) + \sh(2\a\th )
\cth(\a (r^2 +\b^2 ))\right \}J_{\bar z}\ .
\label{solneq}
\end{equation}

The trigonometric solution (\ref{solneq}) has two interesting limits:
rational and constant. Namely, for (\ref{solneq}) in the limit $\a\to 0$, $\b =const$,
we have
\begin{equation}
T_y= \frac{1}{r^2 + \b^2}J_y\ , \quad
T_{\bar y}=\frac{1}{r^2 +\b^2} J_{\bar y} \ ,\quad
T_z=\frac{1}{r^2 +\b^2}J_z \ ,\quad
T_{\bar z}=\frac{1}{r^2 +\b^2 }J_{\bar z}\ ,
\label{solne}
\end{equation}
which solves  the difference Nahm equations (\ref{neq}).
In the limit $\b\to\infty$, $\a = const$, from (\ref{solneq}) we obtain
$$
T_y= 0\ ,\quad 
T_{\bar y}=0\ ,
$$
\begin{equation}
T_z=\frac{1}{2\th }\left\{\ch (2\a\th)-1 +  \sh(2\a\th )\right \}J_z\ ,\
T_{\bar z}=\frac{1}{2\th }\left\{1 - \ch(2\a\th ) + \sh(2\a\th )
\right \}J_{\bar z}
\label{soln}
\end{equation}
as a simplest constant solution of eqs. (\ref{neq}).

In the commutative limit $\th\to 0$, from (\ref{solneq}) we  obtain 
the following solution to the differential Nahm equations (\ref{nah}):
$$
T_y=\frac{\a}{\sh(\a (r^2 + \b^2 ))}J_y\ , \quad
T_{\bar y}= \frac{\a}{\sh(\a (r^2 +\b^2 ))} J_{\bar y} \ ,
$$
\begin{equation}
T_z=\a\, \cth (\a (r^2 +\b^2 )) J_z\ , \quad
T_{\bar z}= \a\, \cth (\a (r^2 +\b^2 ))J_{\bar z}\ .
\label{solnahm}
\end{equation}
In the limit $\a\to 0$, $\b =const$, from (\ref{solnahm}) we obtain  the solution
to (\ref{nah}),
\begin{equation}
T_y=\frac{1}{r^2 + \b^2 }J_y\ , \
T_{\bar y}= \frac{1}{r^2 +\b^2} J_{\bar y} \ ,\
T_z=\frac{1}{r^2 +\b^2} J_z\ , \
T_{\bar z}= \frac{1}{r^2 +\b^2 }J_{\bar z}\ ,
\label{sonahm}
\end{equation}
formally coinciding with the solution (\ref{solne}).
In the limit $\b\to \infty$, $\a =const$, from (\ref{solnahm}) we obtain 
the simplest constant solution
\begin{equation}
T_y=0\ , \quad
T_{\bar y}= 0 \ ,\quad
T_z= \a J_z\ , \quad
T_{\bar z}= \a J_{\bar z}
\label{snahm}
\end{equation}
 to the differential Nahm equations (\ref{nah}).

Above we have described the $u(2)$-valued solutions of the difference Nahm equations (\ref{neq})
and their $\th\to 0$ limit. One may try to find nontrivial abelian solutions as well.
To have explicit $u(n)$-valued solutions with $n>2$, one may take an ansatz (see 
e.g.~\cite{Ivanova:tu, Popov:bf}) reducing Nahm's equations to the finite
Toda lattice equations. In the noncommutative case this will reduce eqs.(\ref{neq})
to a discrete variant of the Toda lattice equations which we will not discuss here.

\medskip

{\bf Solutions of ncSDYM equations.} 
Note that  the difference Nahm equations (\ref{neq}) are integrable
and one can easily obtain the  Lax pair for them by substituting
the ansatz (\ref{ans1}) into the linear system for the ncSDYM equations
written down in~\cite{LP1}.
To obtain the explicit form of noncommutative self-dual gauge potentials in 
$\rct$, one should simply substitute any solution of the difference Nahm
equations (\ref{neq}) into (\ref{ans1}). In particular, by substituting the solution 
(\ref{solneq}) into (\ref{ans1}), one obtains the following solution
of the ncSDYM equation:
$$A_y = - 2\im\bar z \frac{\sh(2\a\th )}{2\th\ \sh(\a (r^2 + \b^2) )}J_y+
 2\im \bar y  \frac{1}{2\th }\left\{1 - \ch(2\a\th ) + \sh(2\a\th )
\cth(\a (r^2 +\b^2 ))\right \}J_{\bar z} \ , 
$$
$$
A_z = - 2\im \bar y \frac{\sh(2\a\th )}{2\th\ \sh(\a (r^2 +\b^2 ))} J_{\bar y}-
 2\im \bar z \frac{1}{2\th }\left\{\ch (2\a\th)-1 +  \sh(2\a\th )
\cth(\a (r^2 +\b^2 ))\right \}J_z     \ , 
$$
$$
A_{\bar y}\  = \  2\im z \frac{\sh(2\a\th )}{2\th\ \sh(\a (r^2 +\b^2 ))} J_{\bar y}\ -\
 2\im  y \frac{1}{2\th }\left\{\ch (2\a\th)-1 +  \sh(2\a\th )
\cth(\a (r^2 +\b^2 ))\right \}J_z \ ,
$$
\begin{equation}
A_{\bar z} = 2\im  y\frac{\sh(2\a\th )}{2\th\ \sh(\a (r^2 + \b^2) )}J_y 
 + 2\im  z \frac{1}{2\th }\left\{1 - \ch(2\a\th ) + \sh(2\a\th )
\cth(\a (r^2 +\b^2 ))\right \}J_{\bar z}\ . \label{nc1}
\end{equation}
In the commutative limit $\th\to 0$, this solution in the coordinates
$(x^\m)\in\rc$ has the form
$$
A_\m^1=2\a\eta^1_{\m\n}\frac{x_\n}{\sh (\a (r^2 + \b^2 ))}\ ,\quad
A_\m^2=2\a\eta^2_{\m\n}\frac{x_\n}{\sh (\a (r^2 + \b^2 ))}\ ,
$$
\begin{equation}\label{pop}
A_\m^3=2\a\eta^3_{\m\n}{x_\n}{\cth(\a (r^2 + \b^2 ))}\ , \
A_\m^4=-\im \, \pa_\m (\ln\ch (\a r^2+\a\b^2))\ .
\end{equation}
 It is the Minkowski solution~\cite{Mink}
of the SDYM equations (\ref{sdym1}) with the nontrivial $SU(2)$ 
components of a gauge potential and the pure gauge extra $U(1)$ piece $A_\m^4$.
These components $A_\m^4$ can be transformed to zero by a {\it complex} gauge 
transformation $A_\m^4\to A_\m^4 + \pa_\m\chi$ with $\chi =\im\ln\ch (\a r^2+\a\b^2)$.
The existence of such a transformation supports the assumption that all 
the above-mentioned solutions of the ncSDYM equations can be transformed to a 
real form by some {\it complex} gauge transformations.

In the limit $\a\to 0$, $\b =const$,  the solution (\ref{solneq}) of the 
difference Nahm equations takes the form  (\ref{solne})
and its substitution into  (\ref{ans1}) gives
$$A_y\ = - 2\im\bar z\frac{1 }{r^2 + \b^2}J_y + 2\im \bar y
 \frac{1 }{r^2 +\b^2 }J_{\bar z} \ , 
\quad
A_z = - 2\im \bar y \frac{1 }{r^2 +\b^2 } J_{\bar y}- 2\im \bar z
  \frac{1 }{r^2 +\b^2 }J_z     \ , 
$$
\begin{equation}\label{nc2}
A_{\bar y}\ = \ 2\im z \frac{1 }{ r^2 +\b^2 } J_{\bar y}\ -\  2\im  y
 \frac{1}{r^2 +\b^2}J_z \ ,
\quad
A_{\bar z}\ =\  2\im  y\frac{ 1}{r^2 + \b^2}J_y 
\ +\ 2\im  z\frac{1 }{r^2 +\b^2}J_{\bar z}\ . 
\end{equation}
It is a solution from~\cite{Correa1}.
In the commutative limit $\th\to 0$, the solution (\ref{nc2}) in the coordinates
$(x^\m)\in\rc$ transforms into the BPST one-instanton solution
\begin{equation}\label{one}
A_\m^a = 2\eta^a_{\m\n}\frac{x_\n}{r^2 + \b^2}
\end{equation}
after removing the extra $U(1)$ pure gauge piece $A_\m^4=-\im \, \pa_\m \ln (r^2+\b^2)$ by a
gauge transformation.

Finally, in  the limit $\b\to\infty$, $\a = const$, from 
(\ref{solneq}) we have obtained the solution (\ref{soln}) to the 
difference Nahm equations. It leads to the following components of the
gauge potential: 
\begin{equation}
A_y = 2\im\a \yb  J_{\zb}\ , \quad A_z = -2\im\a \zb  J_z\ , \quad
A_{\yb} = - 2\im\a  y J_{z} \ , \quad A_\zb = 2\im \a z  J_\zb \ .
\label{anstor}
\end{equation}
This solution is a noncommutative $U(1)$ extension  
\begin{equation}\label{toron}
A_\m^1=0\ ,\quad
A_\m^2=0\ , \quad
A_\m^3=2\a\eta^3_{\m\n}{x_\n}\ ,\quad A_\m^4=-2\im\a{x_\m}\ 
\end{equation}
of the abelian 't~Hooft toron solution~\cite{'tHooft}.
The self-dual Yang-Mills field for the components (\ref{toron}) of 
a gauge potential has constant components
\begin{equation}
F_{y\zb}= F_{\yb z}=F_{yz}=F_{\yb\zb}=0\ , \quad
F_{y\yb}= - 2\a \im J_3 \ , \quad
F_{z\zb}= 2\a \im J_3\ . 
\end{equation}
In the commutative limit, the components $A_\m^4$ corresponding to the second 
$U(1)$ group can be gauge transformed to zero.

\section{Reductions of ncYM equations}

{\bf Noncommutative CFtHW ansatz and ncYM equations.}
In the commutative case, the direct substitution of the CFtHW ansatz~(\ref{cfthw})
into the YM equations (\ref{ym1}) reduces them to the following equation
on a scalar field $\p$:
\begin{equation}\label{mer}
\Box\p + \l\p^3 =0\ ,
\end{equation}
where $\l$ is an arbitrary constant. For $\l =0$, eq.(\ref{mer}) reduces to 
$\Box\p =0$ and its solutions (\ref{laplsol}) describe 't~Hooft $n$-instantons 
 with $n=1,2,...$\ . If $\l\ne 0$, the function  
\begin{equation}\label{me}
\p =\frac{1}{\sqrt{\l r^2}}
\end{equation}
satisfies eq.(\ref{mer}) and provides a non-self-dual solution of the YM equations 
called a {\it meron}~\cite{AFF, Callan}. This solution is singular at $r=0$ and has the
topological charge $Q=1/2$. A two-meron solution can also be obtained by solving 
(\ref{mer})~\cite{AFF, Callan, Actor}. These solutions of the YM equations in $\rc$ 
can be generalized to $\R^{4n}$~\cite{Popov}. Note that such solutions correspond
to unstable $D$0-branes in the type IIB string theory~\cite{Druk}.

The problem of constructing  noncommutative generalizations of meron solutions 
to the YM theory in $\rct$ was not considered yet.
They should exist at least for small values of the noncommutativity parameter 
$\th$ due to the standard deformation theory arguments. In attempting to construct them
we first check the noncommutative generalization (\ref{Amu}) of the  CFtHW ansatz.
After rather lengthy and cumbersome computations we obtain
\begin{align}  
\pa_\m F_{\m y} + [A_\m , F_{\m y} ] & = \nonumber\\[8pt] 
 = -\p^{-\frac{1}{2}}
&\left( \begin{array}{cc} %
\pa_y(\Box\p) - 3 (\pa_y\p)\p^{-1}\Box\p     & 0  \\ 
2\pa_\zb (\Box\p)- 3 (\pa_\zb\p)\p^{-1}\Box\p -3(\Box\p)\p^{-1}\pa_\zb\p   & \quad
-\pa_y(\Box\p) + 3(\Box\p)\p^{-1}\pa_y\p 
\end{array}\right)\p^{-\frac{1}{2}}\ ,
\nonumber
\end{align}
\begin{align}
\pa_\m F_{\m \yb} + [A_\m , F_{\m \yb} ] & = \nonumber\\[8pt] 
 = 
\p^{-\frac{1}{2}}
&\left( \begin{array}{cc} %
\pa_\yb(\Box\p) - 3(\Box\p )\p^{-1} \pa_\yb\p\quad &   
2\pa_z (\Box\p)- 3 (\pa_z\p)\p^{-1}\Box\p -3(\Box\p)\p^{-1}\pa_z\p \\ 0  & 
- \pa_\yb(\Box\p) + 3(\pa_\yb\p)\p^{-1}\Box\p 
\end{array}\right)\p^{-\frac{1}{2}}\ ,
\nonumber
\end{align}
\begin{align}
\pa_\m F_{\m z} + [A_\m , F_{\m z} ] & =
\nonumber\\[8pt] 
= -\p^{-\frac{1}{2}}
&\left( \begin{array}{cc} %
\pa_z(\Box\p) - 3 (\pa_z\p)\p^{-1}\Box\p     & 0  \\ 
2\pa_\yb (\Box\p)- 3 (\pa_\yb\p)\p^{-1}\Box\p -3(\Box\p)\p^{-1}\pa_\yb\p   & 
\quad - \pa_z(\Box\p) + 3(\Box\p)\p^{-1}\pa_z\p 
\end{array}\right)\p^{-\frac{1}{2}}\ ,
\nonumber
\end{align}
\begin{align} 
\pa_\m F_{\m \zb} + [A_\m , F_{\m \zb} ] & =
\label{DF}\\[8pt] 
= \p^{-\frac{1}{2}}
&\left( \begin{array}{cc} %
\pa_\zb(\Box\p) - 3(\Box\p )\p^{-1} \pa_\zb\p\quad &   
2\pa_y (\Box\p)- 3 (\pa_y\p)\p^{-1}\Box\p -3(\Box\p)\p^{-1}\pa_y\p \\ 0  & 
- \pa_\zb(\Box\p) + 3(\pa_\zb\p)\p^{-1}\Box\p 
\end{array}\right)\p^{-\frac{1}{2}}\ .
\nonumber
\end{align}
It is easy to see that the ncYM equations are satisfied if the equations
$$
\pa_y(\Box\p) - 3 (\pa_y\p)\p^{-1}\Box\p = 0\ , \quad 
\pa_y(\Box\p)- 3(\Box\p)\p^{-1}\pa_y\p =0\ ,
$$ 
\begin{equation}\label{**}
\pa_z(\Box\p) - 3 (\pa_z\p)\p^{-1}\Box\p = 0\ , \quad  
\pa_z(\Box\p) - 3(\Box\p)\p^{-1}\pa_z\p  =0\ ,
\end{equation}  
and their hermitian conjugate are satisfied. But from (\ref{**}) it follows that
$$
(\pa_y\p)\p^{-1}\Box\p = (\Box\p)\p^{-1}\pa_y\p \ ,\quad 
(\pa_z\p)\p^{-1}\Box\p = (\Box\p)\p^{-1}\pa_z\p  \ ,
$$ 
\begin{equation}\label{***}
(\pa_\yb\p)\p^{-1}\Box\p = (\Box\p)\p^{-1}\pa_\yb\p \ ,\quad 
(\pa_\zb\p)\p^{-1}\Box\p = (\Box\p)\p^{-1}\pa_\zb\p  \ .
\end{equation} 
{}For $\Box\p\ne 0$ these formulae are not compatible in the noncommutative case, 
since $\pa_\m\p$ does not commute with $\p^{-1}$ and $\Box\p$. In 
the commutative case, however, eqs.(\ref{***}) become identities
and  (\ref{**}) with their complex conjugate become
\begin{equation}\label{m}
\pa_\m (\Box\p + \l\p^3) =0\ .
\end{equation}
Obviously, eqs.(\ref{m}) are satisfied if $\p$ satisfies  eq.(\ref{mer}).

\medskip

{\bf Noncommutative version of the BPST ansatz and merons.}
We see that there are some inconsistency in the reduced ncYM equations obtained
after the substitution of the ncCFtHW ansatz (\ref{Amu}) into (\ref{ncym1}).
Maybe a noncommutative extension of some other modifications of the
CFtHW ansatz (e.g.~\cite{Popov:qs}) will work but this is 
unclear at the moment. Therefore, we shall try to overcome the above difficulties by using
the  ncBPST ansatz (\ref{ans1}) which was successful in the ncSDYM case.
Using the ansatz (\ref{ans2}) and formulae (\ref{ncsdym3}), (\ref{qqq}), we obtain 
\begin{align}
[X_\m , [X_\m , X_y]]\ &=\  -[ X_y , [X_y , X_\yb] + [X_z , X_\zb] ] - 2[X_\zb ,[X_y , X_z]]=
 \nonumber\\[8pt]
 &= -\left\{ 2(\tau -2\th) N^-_\zb Q_1^+ - 2(\tau + 4\th) Q_1^{++}N^+_\zb
- (\tau - 2\th )Q_3 N^+_y  + (\tau + 4\th) N^+_yQ_3^+\right\} \zb -
\nonumber\\[8pt]
  & -  \yb \left\{ 2(\tau + 4\th)N^{++}_yQ_2 - 2(\tau -2\th )Q_2^-N_y - (\tau + 4\th )Q_3^+ N_\zb
+ (\tau - 2\th )N_\zb Q_3\right\}\ ,
\nonumber\\[8pt]
 [X_\m , [X_\m , X_\yb]] &=  [ X_\yb , [X_y , X_\yb ] + [X_z , X_\zb ] ] - 2[X_z,[X_\yb ,X_\zb]]=
\nonumber \\[8pt]
  &= \left\{ 2(\tau -2\th) N^-_\yb Q_1^+ - 2(\tau + 4\th) Q_1^{++}N^+_\yb + (\tau -2\th)Q_3 N^+_z
- (\tau + 4\th)N^+_zQ_3^+\right\} y +
\nonumber\\[8pt]
  & +  z  \left\{ 2(\tau + 4\th)N^{++}_zQ_2 - 2(\tau -2\th)Q_2^-N_z + (\tau + 4\th)Q_3^+ N_\yb
- (\tau - 2\th)N_\yb Q_3\right\}\ ,
\nonumber
\end{align}
\begin{align}
[X_\m , [X_\m , X_z]]\ &=\  - [ X_z , [X_y , X_\yb] + [X_z , X_\zb] ] + 2[X_\yb ,  [X_y , X_z] ]=
 \nonumber\\[8pt]
  &= \left\{ 2(\tau -2\th) N^-_\yb Q_1^+ - 2(\tau + 4\th) Q_1^{++}N^+_\yb + (\tau -2\th)Q_3 N^+_z
- (\tau + 4\th)N^+_zQ_3^+\right\} \zb -
\nonumber\\[8pt]
  & -  \yb  \left\{ 2(\tau + 4\th)N^{++}_zQ_2 - 2(\tau -2\th)Q_2^-N_z + (\tau + 4\th)Q_3^+ N_\yb
- (\tau - 2\th)N_\yb Q_3\right\}\ , 
 \nonumber 
\end{align}
\begin{align}
[X_\m , [X_\m , X_\zb]]\ &=\  [ X_\zb , [X_y , X_\yb] + [X_z , X_\zb] ] + 2[X_y ,[X_\yb , X_\zb]]=
 \label{XXX}\\[8pt]
 &= \left\{ 2(\tau -2\th) N^-_\zb Q_1^+ - 2(\tau + 4\th) Q_1^{++}N^+_\zb
- (\tau - 2\th )Q_3 N^+_y + (\tau + 4\th) N^+_yQ_3^+ \right\} y -
\nonumber\\[8pt]
  & -  z \left\{ 2(\tau + 4\th)N^{++}_yQ_2 - 2(\tau -2\th)Q_2^-N_y - (\tau + 4\th)Q_3^+ N_\zb
+ (\tau - 2\th)N_\zb Q_3\right\}\ .
\nonumber
\end{align}
So, the ncYM equations (\ref{ncym1}) reduce to the system of difference equations
$$
 2(\tau + 4\th)N^{++}_zQ_2 - 2(\tau -2\th)Q_2^-N_z + (\tau + 4\th)Q_3^+ N_\yb
- (\tau - 2\th)N_\yb Q_3 = 0\ ,
$$
$$
 2(\tau -2\th) N^-_\yb Q_1^+ - 2(\tau + 4\th) Q_1^{++}N^+_\yb + (\tau -2\th)Q_3 N^+_z
- (\tau + 4\th)N^+_zQ_3^+ =0\ ,
$$
$$
 2(\tau + 4\th)N^{++}_yQ_2 - 2(\tau -2\th )Q_2^-N_y - (\tau + 4\th )Q_3^+ N_\zb
+ (\tau - 2\th )N_\zb Q_3=0\ ,
$$
\begin{equation}\label{mero}
2(\tau -2\th) N^-_\zb Q_1^+ - 2(\tau + 4\th) Q_1^{++}N^+_\zb  
- (\tau - 2\th )Q_3 N^+_y + (\tau + 4\th) N^+_yQ_3^+ =0\ .
\end{equation}
Recall that $\tau \equiv r^2$ and $N_\m^\pm \equiv N_\m (\tau \pm 2\th ),\ N_\m^{\pm\pm} \equiv 
N_\m (\tau \pm 4\th )$ etc.

It is obvious that for self-dual configurations (satisfying  eqs.(\ref{ncnahm}))
eqs.(\ref{mero}) are also satisfied. 
Moreover, it is not difficult to show that in the commutative limit $\th\to 0$ 
the one-meron solution is reproduced via the ansatz (\ref{ans1}) for
\begin{equation}
T_y=\frac{2\im}{\sqrt{\l r^2}}J_y\ ,\quad T_\yb=\frac{2\im}{\sqrt{\l r^2}}J_\yb\ ,\quad
T_z=\frac{2\im}{\sqrt{\l r^2}}J_z\ ,\quad T_\zb=\frac{2\im}{\sqrt{\l r^2}}J_\zb\ .
\end{equation}
To find meron-type solutions of the ncYM equations, one should obtain nontrivial
solutions of the system of difference equations (\ref{mero}).
We postpone this task for a future work.

\section{Concluding remarks}

In this paper  we have discussed the noncommutative CFtHW-like 
ansatz~\cite{Correa1, LP1} for a gauge potential 
 and showed that it reduces the ncYM equations to the system (\ref{**}) of equations
on a scalar field $\phi$. Generically, in the non-self-dual case this system is not compatible.
To resolve this problem we generalized the BPST-like ansatz from the paper~\cite{Correa1}
and derived the system (\ref{mero}) of difference equations by substituting this ansatz into the
ncYM equations. For the self-dual subcase, the above-mentioned system reduces to 
the difference Nahm equations. We obtained some new solutions of the ncSDYM 
equations via solving the difference Nahm equations.

It will also be interesting to consider solutions of the SDYM equations 
on a noncommutative version of pseudo-Euclidean space with a metric of signature
$(2,2)$, since these equations appear in a zero-slope limit of $N=2$ strings
 with a nonzero constant $B$-field~\cite{Le:2000nm}. Moreover,
nonlocal dressing symmetries of the self-dual Yang-Mills~\cite{6}
raise to (dressing) symmetries of open $N=2$ strings~\cite{5} 
that helps to construct solutions of Berkovits' string field theory~\cite{Berkovits:1995ab}
via the dressing approach~\cite{LPU}. Therefore,
constructing solutions of the (noncommutative) SDYM theory may be helpful for
finding solutions in string field theory and studying its nonperturbative properties. 
Recall that the (noncommutative) SDYM equations are integrable. At the same time
 the (noncommutative) YM equations are not integrable and  there are no general methods
of constructing their solutions. Therefore, some guesswork with  appropriate
ans\"atze may be useful in producing interesting solutions to the ncYM equations.

\bigskip

\noindent
{\large{\bf Acknowledgements}}

\smallskip
\noindent
The authors are grateful to Alexander Popov for reading  the manuscript 
and useful discussions. F.F.-S. thanks the Albert-Ludwig-Fraas-Stiftung for 
financial support, 
the work of T.A.I. was partially supported by the Heisenberg-Landau Program.
The authors acknowledge the hospitality of the Institut f\"ur Theoretische Physik, 
Universit\"at Hannover where this work was done within the framework 
of the DFG priority program (SPP 1096) in string theory.

\end{document}